# Large and Persistent Photoconductivity due to Hole-Hole Correlation in CdS

Han Yin[1], Austin Akey[2], R. Jaramillo[3]

[1] Department of Mechanical Engineering, Massachusetts Institute of Technology, Cambridge, MA, 02139 USA
[2] Center for Nanoscale Systems, Harvard University, Cambridge, MA 02138 USA
[3] Department of Materials Science and Engineering, Massachusetts Institute of Technology, Cambridge, MA, 02139 USA

Large and persistent photoconductivity (LPPC) in semiconductors is due to the trapping of photo-generated minority carriers at crystal defects. Theory has suggested that anion vacancies in II-VI semiconductors are responsible for LPPC due to negative-$U$ behavior, whereby two minority carriers become kinetically trapped by lattice relaxation following photo-excitation. By performing a detailed analysis of photoconductivity in CdS, we provide experimental support for this negative-$U$ model of LPPC. We also show that LPPC is correlated with sulfur deficiency. We use this understanding to vary the photoconductivity of CdS films over nine orders of magnitude, and vary the LPPC characteristic decay time from seconds to $10^4$ seconds, by controlling the activities of $Cd^{2+}$ and $S^{2-}$ ions during chemical bath deposition. We suggest a screening method to identify other materials with long-lived, non-equilibrium, photo-excited states based on the results of ground-state calculations of atomic rearrangements following defect redox reactions, with a conceptual connection to polaron formation.

## 1. Introduction: Persistent photoconductivity, defect models, and lattice relaxation

Large and persistent photoconductivity (LPPC) - wherein the photoconductive response of a semiconductor is enormous and can persist for many hours after illumination is turned off - is associated with crystalline defects. LPPC has been reported in many semiconductors including Si, III-Vs, oxides and chalcogenides [1–9]. LPPC in cadmium sulfide (CdS) has long been reported, and is relevant to the operation of thin-film solar cells that use CdS as an $n$-type layer [10–12]. Here, we show that chemical bath deposition (CBD) can be used to tune the photoconductivity of CdS films over nine orders of magnitude, and vary the photoconductivity decay time from seconds to $10^4$ seconds. We vary the activities of $Cd^{2+}$ and $S^{2-}$ ions in the chemical bath and demonstrate that LPPC results from sulfur deficiency in CdS. We provide experimental support for the theoretical result that LPPC is caused by so-called negative-$U$ sulfur vacancies, at which charge-lattice coupling results in an effective attractive force between positively-charged holes. Due to this lattice distortion, recombination becomes thermally-activated, slowing the return to equilibrium. This is conceptually related to the "DX-center" lattice distortion responsible for dopant de-activation in AlGaAs, and to long photo-excitation lifetimes observed in systems with polaron excited states such as organic semiconductors and halide perovskites [13–17]. The idea that atoms rearrange to create physical separation between photo-excited charge carriers, thus slowing the decay to equilibrium, is also reminiscent of biological light absorbers such as photosystem II and dye molecules used for solar energy conversion [18,19].

Photoconductivity in large band gap II-VI materials is due to majority-carrier transport while the photo-generated minority carriers are trapped at crystal defects. Large and persistent photo-effects are associated with atomic lattice distortions around defects in response to changes in electron occupation, for instance at DX centers in AlGaAs, and at wrong-coordinated sites in amorphous Se [20,21]. Zhang, Wei, Lany, and Zunger (ZWLZ) published theoretical proposals that LPPC in II-VI and chalcopyrite semiconductors is due to lattice relaxation at anion vacancies [22,23]. Anion vacancies ($V_{An}$) in equilibrium in $n$-type semiconductors tend to be neutrally-charged, deep donors (e.g. $V_O^\times$ in ZnO). In CdS at equilibrium, the cations surrounding $V_S^\times$ distort inwards, as they are attracted by the two un-bound electrons and no longer repelled by a sulfur ion. Upon photo-excitation, anion vacancies can release two electrons and become doubly-charged ($V_{An}^{\bullet\bullet}$). Without subsequent lattice relaxation, the positively-charged, deep donors would quickly capture electrons from the conduction band. However, $V_{An}^{\bullet\bullet}$ sites no longer contain electrons to screen cation-cation interactions: the cations repel each other and distort outwards. We denote this transition to a non-equilibrium, metastable configuration as $V_{An}^{\bullet\bullet} \rightarrow (V_{An}^{\bullet\bullet})^* + ph$, where $ph$ indicates interactions with lattice vibrations. The resulting, distorted lattice presents an activation barrier to re-capture of photoexcited electrons by $(V_{An}^{\bullet\bullet})^*$. The distorted lattice may even raise the $(V_{An}^{\bullet\bullet})^* \rightarrow (V_{An}^{\bullet})^*$ and $(V_{An}^{\bullet})^* \rightarrow (V_{An}^{\times})^*$ charge transition levels above the conduction band edge ($E_C$), further reducing the rate of electron re-capture. This may be the case for ZnO, for which theory finds that the $(V_{An}^{\bullet\bullet})^* \rightarrow (V_{An}^{\bullet})^*$ transition level is resonant with the conduction band [23].

In **Fig. 1** we visualize the distortion around $V_S$ in the $V_S^\times$ and $(V_S^{\bullet\bullet})^*$ charge states, as calculated by Nishidate *et al.* [24]. These predictions using density functional theory are consistent with the above description of contraction and expansion of $V_S$-$Cd_4$ tetrahedra in response to photo-excitation.

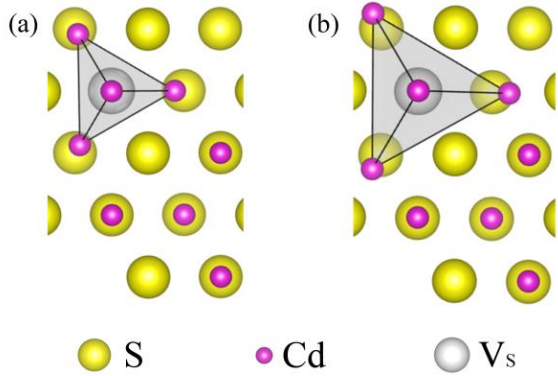

**Figure 1:** Illustration of lattice distortions around $V_S$ in the $V_S^{\times}$ ground state (a) and $(V_S^{\bullet\bullet})^*$ non-equilibrium state (b). The bond lengths are as calculated by Nishidate *et al.* [24]. The $V_S$ site is indicated by a grey sphere. The orthographic projection along a Cd-S bond allows easy comparison of the distortion of the Cd$_4$ tetrahedron in the undistorted lattice (bottom-right of each panel) and around the vacancy.

## 2. Experimental procedures and results

We make CdS thin films on glass substrates by chemical bath deposition (CBD) [25,26]. The bath contains a cadmium source, a sulfur source such as thiourea (SC(NH$_2$)$_2$), and a complexing agent. One proposed reaction mechanism is as follows [27]:

1. $Cd^{2+} + 4NH_3 \leftrightarrow Cd(NH_3)_4^{2+}$             (1)
2. $SC(NH_2)_2 + OH^- \leftrightarrow CH_2N_2 + H_2O + SH^-$
3. $SH^- + OH^- \leftrightarrow S^{2-} + H_2O$
4. $S^{2-} + Cd(NH_3)_4^{2+} \leftrightarrow CdS + 4NH_3$

Chemical equilibria in the bath affect the composition of the growing CdS films. Therefore, CBD offers a convenient way to control the composition of the resulting films by adjusting the bath chemistry.

Our film growth is consistent with an initial ion-by-ion process followed by a cluster-by-cluster process (see Supplemental Material [28]). We measure film composition using X-ray fluorescence (XRF) data, which we calibrate using wavelength diffraction spectroscopy (WDS) measurements on select samples. We selected one sample each with high (10$^7$) and no photoresponse for high-accuracy WDS measurements, and then we linearly scaled the XRF results according to this two-point calibration. X-ray diffraction (XRD) reveals that the films are polycrystalline with the dominant phase being cubic sphalerite, as shown in **Fig. 2a**. The broad peak at 20-35° is due to the glass substrates. The peaks show a slight shift of about 0.5° higher than the reference pattern, probably indicating compressive strain. Grain sizes obtained using Scherrer's method range from 40 to 50 nm, and increase with deposition time. Transmission electron microscopy (TEM) (**Fig. 2b**) shows that the thin film is well-crystalized with no evidence for amorphous regions. We used optical reflection and transmission measurements to estimate the band gap ($E_g$) using Tauc's method. For films thicker than approximately 50 nm we find $E_g = 2.45 \pm 0.08$ eV. For thinner films, $E_g$ tends to increase and the data is difficult to analyze accurately due to multiple film-substrate reflections [28]. We measured Hall mobility $\mu_H = 1.57 \pm 0.27$ cm$^2$ V$^{-1}$ s$^{-1}$ for a typical sample using a rotating parallel dipole line Hall system with the film under white-light illumination [29]. In the analysis that follows, we assume that the drift mobility is equal to $\mu_H$ and is independent of carrier concentration.

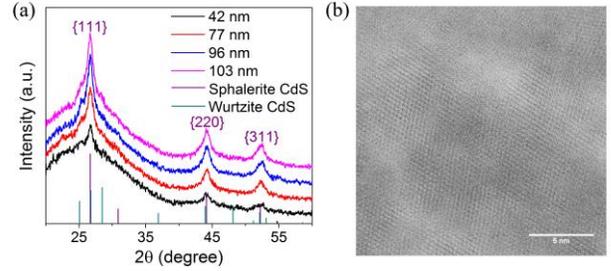

**Figure 2:** Structural characterization of CdS thin films. (a) XRD spectra measured in grazing incidence geometry for CdS films (CdS_160709_5-8) grown at 70 °C with different thickness for a fixed bath chemistry (0.0015 M cadmium nitrate, 0.075 M thiourea and 1.87 M ammonia). Purple lines and labeled peaks correspond to the cubic sphalerite phase (ICDD #96-900-0109), and green lines to the hexagonal wurtzite phase (ICDD #04-004-8895). (b) TEM micrograph of a representative sample (CdS_160704_2) shows that it is well-crystallized.

We measure photoconductivity in two-point configuration using silver paint contacts on bare CdS films and an electrometer (Keithley 6517B). We estimate carrier concentration ($n$) from conductivity ($\sigma$) using the relation $n = \frac{\sigma}{e\mu_H}$. For the data reported here, we held samples in the dark until steady-state conductivity was reached, and then switched on illumination for three hours. The measurement probe station is surrounded by a dark enclosure. The illumination sources used for this study are an AM1.5 solar simulator, a cold white LED (Thorlabs Solis 1A), and a 455 nm LED (Thorlabs M455L3). We used a thermal chuck to control sample temperature.

To understand and control photoconductivity in CdS thin films, we study the effect of changing the bath chemistry. The bath composition was varied around a baseline chemistry, *i.e.* 0.0015 M cadmium source, 0.075 M thiourea and 1.87 M ammonia. In **Fig. 3a-c** we show the effect of varying the concentration of cadmium, thiourea, and ammonia on the phoconductivity magnitude and decay. For this chemical study we charactize photoconductivity by two metrics: photosensitivity and decay rate. Photosensitivity is the ratio of conductivity under illumination to that in the dark, and the decay rate is obtained by fitting a single exponential function to the decay curve. The full decay dynamics are more complex than can be captured by a single time constant (*cf.* **Fig. 5a**), but a simplification suffices here. For high cadmium concentration, films exhibit LPPC. The trend is reversed with thiourea (TU) concentration: for high TU concentration, films are not photosensitive (and therefore the decay rate trends towards zero). Ammonia has a similar effect as TU: higher concentrations of ammonia weaken photoconductivity. The role of ammonia in CBD is twofold: to complex with free $Cd^{2+}$ ions, and to accelerate the hydrolysis of TU, as illustrated by the following reactions:

$$Cd^{2+} + 4NH_3 \leftrightarrow Cd(NH_3)_4^{2+} \qquad (2)$$
$$SC(NH_2)_2 + OH^- \leftrightarrow CH_2N_2 + H_2O + SH^-$$

As a result, a higher concentration of $NH_3$ reduces the concentration of free $Cd^{2+}$ and increases the concentration of sulfur ions. Therefore, the effect of $NH_3$ on photoconductivity is consistent with the previous two cases: LPPC is enhanced by cadmium-rich, sulfur-poor bath chemistry. Other variables that affect photoconductivity include the concentration of ammonium ($NH_4^{+}$) ions, the choice of cadmium salt, the bath stirring speed, and the bath temperature [28].

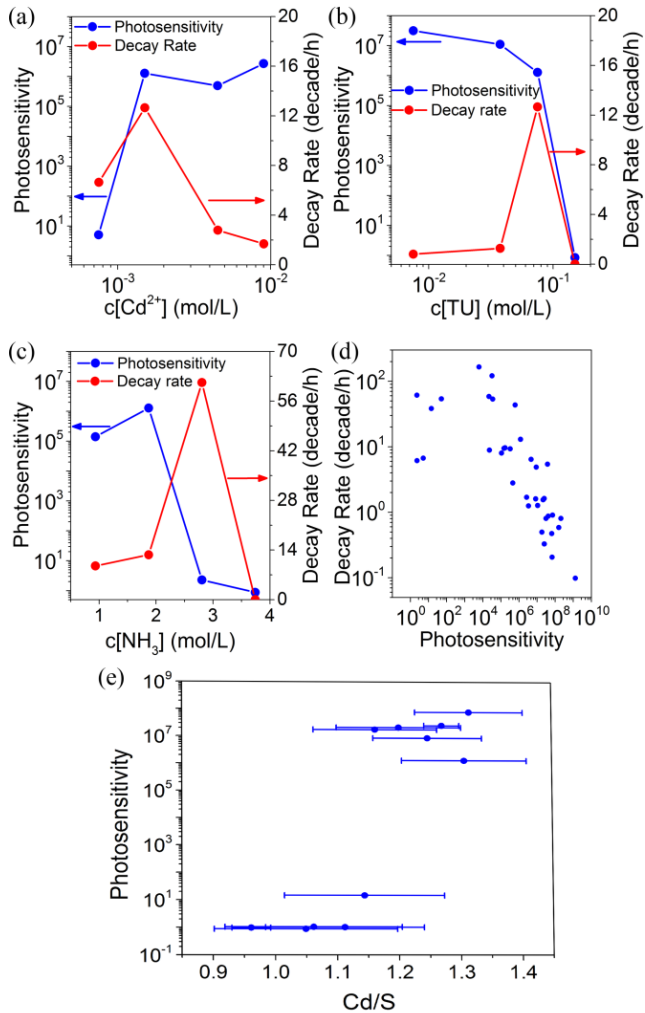

**Figure 3:** Variation of photoconductivity sensitivity and decay rate with CBD bath chemistry using AM1.5 illumination. (a) Varying cadmium source concentration. (b) Varying thiourea (TU) concentration. (c) Varying ammonia concentration. These and other studies point to the ratio of cadmium-to-sulfur activity in the bath as the determining factor for photoconductivity. Note that a decay rate approaching zero can result when the film is not photosensitive, *e.g.* for high [TU]. (d) Decay rate and sensitivity are negatively correlated for photosensitive films. (e) Relationship between photosensitivity and Cd/S ratio in CdS films. Errorbars represent 68% confidence intervals from the counting statistics of the XRF measurements.

Since sulfur-deficient chemical baths produce CdS films with pronounced photoconductivity, we hypothesize that highly-photosensitive films are sulfur-deficient. Our film composition measurements by XRF and WDS support this hypothesis. We show **Fig. 3e** that photosensitivity is strongly correlated with sulfur deficiency. This trend is also confirmed by Rutherford backscattering spectroscopy (RBS) measurements. We performed RBS on the same two samples used for WDS calibration. The Cd/S ratios are $1.25 \pm 0.04$ for the photosensitive sample and $0.91 \pm 0.04$ for the one without photoresponse, compared with $1.27 \pm 0.03$ and $0.96 \pm 0.03$ measured by WDS. The Cd/S ratios in **Fig. 3e** span a range from 0.96 to 1.31, which would correspond to enormous point defect concentrations (up to $5\times10^{21}$ cm$^{-3}$). The transition between strongly- and weakly-photosensitive samples takes place over a narrow range of less than ±2%. This suggests that intrinsic point defects are primarily responsible for LPPC, and that the concentration of relevant point defects saturates for larger deviations in overall composition. Inspired by previous studies connecting anion vacancies to LPPC in other materials [23,30,31], we propose that sulfur vacancies are responsible for large and persistent photoconductivity in CdS.

Knowing the chemical cause of LPPC allows us to engineer it. By systematically varying bath chemistry and temperature, we can make films with sensitivity of up to $10^9$ and decay rate of 0.1 decade/h [28].

## 3. Modeling photoconductivity
### 3.1. The standard model

Boer and Vogel described majority-carrier photoconductivity (*n*-type discussed here) using a generic model that considered two defects with charge transition energy in the band gap: a recombination level and an electron trap level [32]. Electrons can be optically excited from the recombination level to the conduction band at a rate of $F$, and re-captured with rate coefficient $R_r$. The trap level has emission and capture rate coefficients $P$ and $R_t$, respectively. The rate equations are:

$$\frac{dn}{dt} = F + Pn_t - R_t(N_t - n_t)n - R_r(n + n_t)n \qquad (3)$$
$$\frac{dn_t}{dt} = R_t(N_t - n_t)n - Pn_t \qquad (4)$$

where $n$ is the density of conduction band electrons, $n_t$ the density of trapped electrons, and $N_t$ the total trap density. The model assumes that all conduction band electrons are generated from the recombination level, and that the concentration of holes in the valence band is negligible. The recombination level is typically a deep donor, and is sometimes called the sensitizing level [33]. Under illumination, electrons are effectively transferred from deep donors to shallow traps. When illumination is turned off, relaxation to equilibrium is often limited by the trap escape rate. In the following discussion, we will refer to this as the standard model for photoconductivity.

To apply the standard model to CdS, we hypothesize that the deep donors are sulfur vacancies. The traps could include several species, possibly including ionized sulfur vacancies. There will be band-to-band generation, and we assume that the capture of free holes by recombination levels is effectively instantaneous. In the following we discuss our experimental results in light of the standard model, suggesting modifications as needed to describe the data.

*3.2 Spectral dependence of photoconductivity*

In **Fig. 4** we show the dependence on photoconductivity on excitation wavelength. For this experiment the sample was equilibrated in the dark at room temperature, and then monochromatic incident light was varied from long-to-short wavelength (λ). The light was modulated using a mechanical chopper at 31 Hz, a constant voltage was applied across the sample, the current response was amplified and de-modulated, and the resulting response curve was normalized by the wavelength-dependence of the incident light power. The data show a flat response and are noise-limited above 950 nm (below 1.3 eV), a rising trend for λ < 950 nm, and a steep rise near $E_g$. The broad photoconductive response for 600 < λ < 950 nm (1.3 – 2 eV) indicates the energy inside the band gap of deep donors ("sensitizing centers") that contribute to photoconductivity.

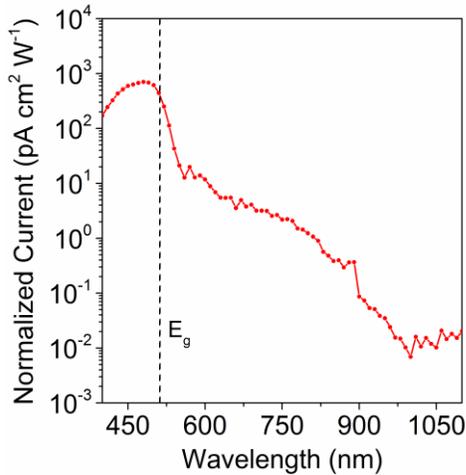

**Figure 4:** Photoconductivity excitation spectra for a representative sample with large photoresponse (CdS_160613_4).

*3.3. Modeling photoconductivity decay data*

In **Fig. 5a** we show a typical photoconductivity timeseries measured at room temperature. The photoconductivity decay after illumination is turned off has three distinct regimes (1) an initial, rapid decay, (2) an intermediate, nearly single-exponential decay, (3) a long-time decay, slower than exponential. The short-time decay is well-modeled by a stretched-exponential:

$$\sigma_{PC}(t) = \sigma_{PC,0} \exp\left(-\left(\frac{t}{\tau_{str}}\right)^\beta\right) \quad (6)$$

$\sigma_{PC}(t)$ and $\sigma_{PC,0}$ are the conductivity due to excess carriers at times $t$ and time $t = 0$, respectively. $\sigma_{PC}$ is defined as $\sigma(t) - \sigma_s$, where $\sigma_s$ is the conductivity in equilibrium in the dark. We approximate $\sigma_s$ as the steady-state conductivity measured prior to illumination, after a long time held in the dark. $\tau_{str}$ and $\beta$ are fitting parameters corresponding to a characteristic decay time and the width of the decay time distribution, respectively. In **Fig. 5b** we show fits of Eqn. 6 to decay data at short times for different measurement temperatures. The stretched exponential function is the Laplace transform of function $P(s, \beta)$:

$$\exp\left(-\left(\frac{t}{\tau_{str}}\right)^\beta\right) = \int_0^\infty P(s, \beta) \exp\left(-s\frac{t}{\tau_{str}}\right) ds \quad (7)$$

$P(s, \beta)$ describes a distribution of single-electron exponential decay processes with decay times $\tau = \frac{\tau_{str}}{s}$. The distribution of $\beta$ at different temperatures is 0.43 ± 0.01. This narrow distribution means that the activation energy distribution is unchanged within the measured temperature range. For a thermally-activated process, $\tau$ is related to an activation energy $E_a$ through the Arrhenius equation $\tau = \tau_0 \exp\left(-\frac{E_a}{kT}\right)$. Our temperature-dependent data (**Fig. 5b**) show that photoconductivity decay is thermally-activated with $\tau_0 = 15 \pm 12$ μs.

The intermediate-time decay is well-modelled by a single exponential. This can be shown explicitly by fitting the data to a stretched exponential (**Eqn. 6-7**): for this fit β = 0.9960 ± 0.0003 and the distribution of activation energy is sharply peaked.

In **Fig. 5d** we plot the distribution of activation energy photoconductivity decay. The distribution for short-time decay does not change with temperature over the measured range of 25 - 105 °C, which is reasonable because relatively small changes should not affect the distribution of defect levels or the phonon spectra. The intermediate-time decay process has an activation energy of $E_a$ = 0.55 eV, which falls at the high end of the distribution of activation energy for the short-time decay. This is therefore the slowest single-electron process and remains active after the faster processes with lower activation energies have completed.

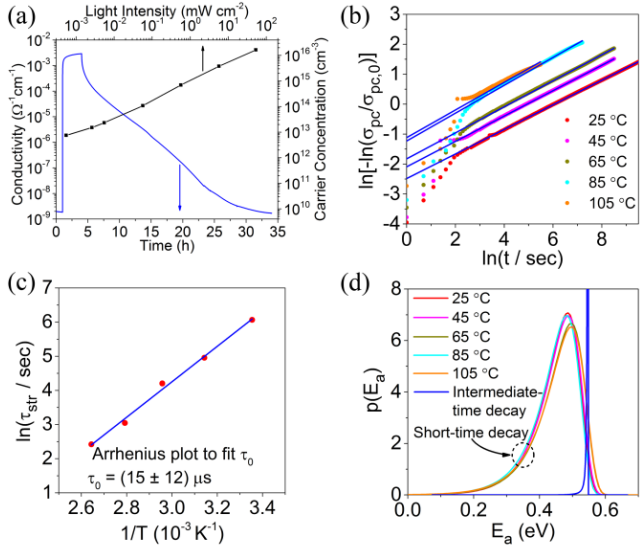

**Figure 5:** Analysis of photoconductivity decay for a representative sample with large photoresponse (CdS_160613_4). (a) Bottom axis, blue curve: photoconductivity time-series data measured using a white LED, which was turned on at time t = 1 h and off at time t = 4 h. Top axis, black curve: power law dependence of steady-state photoconductivity on light intensity, measured using a 455 nm LED. (b) Stretched exponential fits (blue lines) to short-time decay data measured at different temperatures. (c) Arrhenius plot; red points are data, blue line is Arrhenius fit. (d) Probability distribution (p($E_a$)) of activation energy for short- and intermediate-time decay processes.

The photoconductivity decay dynamics can be understood in the context of the standard model. The rapid decay at short-times is due to trapping and recombination of electrons in the conduction band. The relative magnitude of recombination and trapping rate depends on temperature [28]. The single-exponential decay at intermediate times implies $n_t \gg n$: most of the non-equilibrium electrons are trapped. In this case $\frac{dn_t}{dt} \approx 0$ and **Eqn. 3** becomes $\frac{dn}{dt} = -R_r n_t n$, which produces single-exponential decay. This approximation relies on recombination being slow, so that $n_t$ changes slowly. These approximations are supported by numerical simulation of the standard model [28]. At long times the system approaches equilibrium, we have detailed balance between the trap level and conduction band. **Eqn. 4** gives $n_t = \frac{R_t N_t}{P} n$ considering $n_t \ll N_t$. Then **Eqn. 3** becomes $\frac{dn}{dt} = -R_r \left(\frac{R_t N_t}{P} + 1\right) n^2$ and the decay rate is non-linear [28].

### 3.4. Power-law model of steady-state photoconductivity

In **Figs. 5a and 6** we show the dependence of steady-state photoconductivity on light intensity. The data can be divided into two regimes: under low illumination it follows a power law $\sigma_{PC} \propto I^b$ with exponent $b$ between 0.3 and 0.5 that depends on temperature, while under higher illumination b is between 0.6 and 0.7.

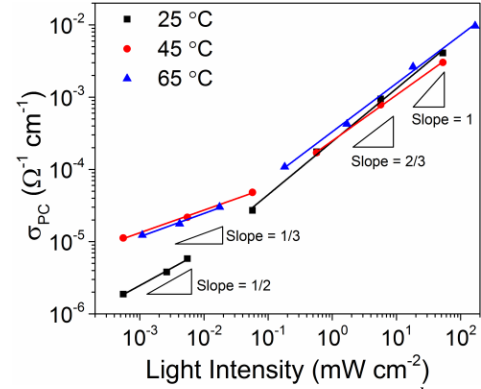

**Figure 6:** The power-law relationship $\sigma_{PC} \propto I^b$ between the steady-state photoconductivity ($\sigma_{PC}$) and light intensity (*I*) measured at different temperatures for a representative sample with large photoresponse (CdS_160613_4). The data at 25 °C (black curve) is the same as plotted in **Fig. 5a**. At low intensity *b* is between 0.3 and 0.5. At higher illumination *b* is between 0.6 and 0.7 Data was measured using a 455 nm LED.

We can model the generation rate as $F = g\Phi$, where $\Phi$ is photon flux and $g$ is a constant. Adding this term to the standard model and setting time derivatives to zero yields the steady-state condition $g\Phi = R_r(n + n_t)n$. For high trap occupancy $n_t \approx N_t \gg n$, the standard model predicts a power law $n \propto \Phi$. For the case of low trap occupancy $n_t \ll N_t$, setting **Eqn. 4** to zero produces $n_t = \frac{R_t N_t}{P} n$. In this case $g\Phi = R_r \left(\frac{R_t N_t}{P} + 1\right) n^2$, which leads to $n \propto \sqrt{\Phi}$. Therefore, the standard model with linear recombination can produce an apparent power law with exponent $0.5 < b < 1$, approaching 1 under high illumination. In order to model our results showing $1/3 < b < 2/3$, we consider the effect of quadratic recombination. If we replace the linear recombination term by $R'_r(n + n_t)n^2$, the power law relationships becomes $n \propto \sqrt{\Phi}$ when $n_t \approx N_t$, and $n \propto \Phi^{\frac{1}{3}}$ when $n_t \ll N_t$. Our data is therefore consistent with the combined effect of linear and quadratic recombination, yielding an apparent exponent 1/3 < b < ½ at low illumination, and ½ < b < 1 at high illumination. We suggest that the exponent approaching 1/3 at low illumination becomes more pronounced at high temperature because the quadratic recombination process is thermally-activated.

### 3.5. Temperature-dependence of steady-state photoconductivity

In **Fig. 7a** we plot the temperature-dependence of the steady-state conductivity in the dark and under illumination. At equilibrium in the dark, the conductivity increases with increasing temperature. This is as-expected for a non-degenerate and large band gap semiconductor. In contrast, the photoconductivity decreases with increasing temperature. In the context of the standard model, this shows that recombination is thermally-activated.

Thermally-activated recombination may have several causes. By comparing energy scales, we can rule out thermal release of holes trapped at sensitizing levels. The spectral photoconductivity data in **Fig. 4** show that the sensitizing levels are in a band approximately 1.3 – 2 eV below the conduction band edge, or 0.4 – 1.1 eV above the valence band edge. The activation energy determined from the steady-state photoconductivity data in **Fig. 7a** is 0.144 ± 0.024 eV. Therefore, thermal release of holes trapped at the same energy levels as the sensitizing levels cannot explain the observations.

*3.6. Temperature-dependence of photoconductivity transients*

In **Fig. 7b** we show photoconductivity transient data measured at different temperatures. The transient behavior under illumination can be described by two time scales: a fast rise with time constant $\tau_1$, and a slow decay with time constant $\tau_2$. Photoconductivity decay with the light off is characterized by a third time scale, $\tau_3$. The time constants $\tau_1$ and $\tau_3$ appear to be correlated: both get smaller with increasing temperature. The correlation between $\tau_1$ and $\tau_3$ is more shown in **Fig. 7c**. The best-fit slope is 0.94 ± 0.15 and the p-value is 0.004 for a one-tailed t-test, which means that the slope is greater than 0 at 95% confidence level. That the slope is nearly unity means that $\tau_1$ and $\tau_3$ evolve similarly with temperature. This suggests that the same thermally-activated process is responsible for these two transients.

The combination a fast rise (with time constant $\tau_1$) and a slow decay (with time constant $\tau_2$) leads to a transient maximum (an "overshoot") in photoconductivity under illumination. This maximum is visible in the data in **Fig. 7b** measured between 378 and 338 K; at lower temperatures, the maximum moves outside our time window. We find that $R_r > R_t$ is a necessary condition for the rate equations to allow a transient maximum ( $dn/dt = 0$ ) under illumination [28]. The observation that the maximum occurs at shorter times with increasing temperature means that recombination is thermally-activated, and grows faster than $R_t$ as temperature increases. A similar conclusion can be drawn when considering quadratic recombination [28].

In **Fig. 7d** we show the results of numerical simulation of the standard model with thermally-activated recombination, i.e. $R_r = R_{r0} \exp\left(-\frac{E_a}{kT}\right)$. For simplicity, in this simulation we assume that trapping rate is temperature-independent, because our data do not clearly indicate the temperature dependence of the trapping rate (adding thermally-activated trapping leads to the same qualitative conclusions). The simulations reproduce the conductivity overshoot and its temperature dependence, and the positive correlation between time constants $\tau_1$ and $\tau_3$. From the data and simulations in **Fig. 7** we conclude that thermally-activated recombination is responsible for the temperature dependence of LPPC in CdS.

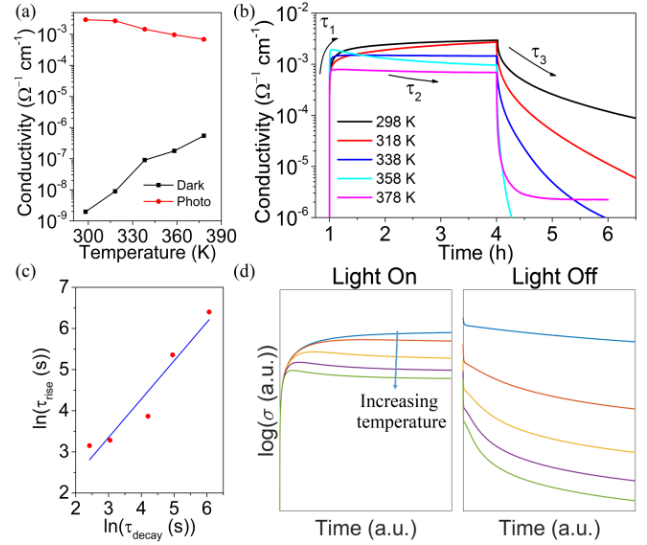

**Figure 7:** Evidence for thermally-activated recombination in the temperature-dependence of photoconductivity for a representative sample with large photoresponse (CdS_160613_4). (a) Temperature dependence of steady-state conductivity in the dark and under illumination using the white LED. (b) Transient photoconductivity measured at different temperatures. The light is switched on at 1 hr and off at 4 h. The transient under illumination shows a fast rise with time constant $\tau_1$ and a slow decay with time constant $\tau_2$. The decay with the illumination off has characteristic time constant $\tau_3$. (c) Positive correlation between time constants $\tau_1$ and $\tau_3$. (d) Numerical simulations of the standard model, showing how transient behavior changes with temperature.

**4. Discussion**

*4.1. Modeling large and persistent photoconductivity in CdS*

The standard model of photoconductivity is simple but possess high explanatory power. In **Sec. 3** we analyzed our results on CdS in terms of the standard model, and suggested modifications needed to describe the data. Both linear and quadratic recombination processes are needed to describe the observed power law dependence of steady-state photoconductivity on illumination. The temperature dependence of steady-state photoconductivity and of the transient maximum observed under illumination indicate that recombination of conduction electrons with trapped holes is thermally-activated.

In **Sec. 2** we showed that photosensitivity is strongly correlated with sulfur deficiency (**Fig. 3e**). We also showed that the photosensitivity and photoconductivity decay rate are strongly correlated (**Fig. 3d**). Therefore, the same defects that are responsible for the large photoconductive response are also responsible for the slow decay. This is supported by our simulations showing that recombination is the rate-limiting process during photoconductive decay. This is unlike the usual interpretation of the standard model in which the size of the steady-state photoresponse depends on the sensitizing centers, while the slow decay depends on the trap levels.

These observations support the hypothesis that sulfur vacancies featuring hole-hole correlation with negative-$U$ behavior are responsible for large photoconductivity and its slow decay in CdS. Both the large photoresponse and its slow decay originate from thermally-activated recombination between conduction electrons and trapped holes.

The resulting, modified standard model to describe LPPC in CdS is consistent with the ZWLZ model developed to explain LPPC in ZnO (**Sec. 1**). The sensitizing levels are neutral sulfur vacancies, $V_S^\times$. The model features thermally-activated linear and quadratic recombination processes, *i.e.* $(V_S^{\bullet\bullet})^* \to V_S^\bullet$, $(V_S^\bullet)^* \to V_S^\times$, and $(V_S^{\bullet\bullet})^* \to V_S^\times$. Each of these processes involves a transition between metastable and equilibrium lattice distortions around $V_S$. The kinetic pathway for quadratic recombination could be (1) Thermally-assisted, two-electron excitation into the vacancy level, followed by spontaneous lattice relaxation: $(V_S^{\bullet\bullet})^* + 2e^- \to (V_S^\times)^* \to V_S^\times + ph$; or, (2) Thermally-assisted lattice distortion, followed by two-electron capture: $(V_S^{\bullet\bullet})^* + ph \to V_S^{\bullet\bullet}$, $V_S^{\bullet\bullet} + 2e^- \to V_S^\times$. Pathways for linear recombination involving a metastable state $(V_S^\bullet)^*$ can be proposed in a similar way. The chemical identify of the traps is unknown. The population of traps may include singly-ionized sulfur vacancies with or without a lattice distortion, $V_S^\bullet$ or $(V_S^\bullet)^*$ in our notation.

### 4.2. Energy level diagram

In **Fig. 8a** we show a generic energy level diagram for the standard model of photoconductivity. Energy level diagrams show equilibrium charge transition levels and do not clearly represent kinetic effects such as thermally-activated recombination and metastable lattice distortions. Nevertheless, we can illustrate features of the energy level diagram for photoconductive CdS based on our data.

The Fermi level ($E_F$) at thermal equilibrium in the dark can be calculated using the formula $n_0 = N_C \exp\left(-\frac{E_C - E_F}{kT}\right)$, where $n_0$ is the carrier density in the dark, and $N_C$ is the conduction band effective density of states, $N_C = 2.2 \times 10^{18}$ cm$^{-3}$ [34]. Our Hall measurements give $n_0 = (7.0 \pm 1.0) \times 10^9 cm^{-3}$ at room temperature (298 K), from which we calculate $E_C - E_F = 0.507 \pm 0.003$ eV. Similarly, our Hall data give $n_{PC,0} = (1.0 \pm 0.1) \times 10^{16} cm^{-3}$ and $E_C - E_F = 0.140 \pm 0.003$ eV under 149 mW/cm$^2$ white LED illumination. The photoconductivity excitation spectrum gives $E_C - E_R > 1.3$ eV, where $E_R$ is the charge transition level of deep donors in equilibrium in the dark.

In **Fig. 8b** we show the energy level diagram in the dark. $E_F$ lies in between the deep donors (including $V_S^\times$) and the shallow traps. In **Fig. 8c** we show one possibility for the energy level diagram under illumination with sulfur vacancies in their ionized, metastable configuration $(V_S^{\bullet\bullet})^*$. Here we have drawn the transition level $(V_S^{\bullet\bullet})^* \to (V_S^\times)^*$ resonant with the conduction band, as for ZnO, but this is speculative. The conclusion that the dominant recombination process is thermally-activated does not depend on the location of the charge transition levels relative to the band edges.

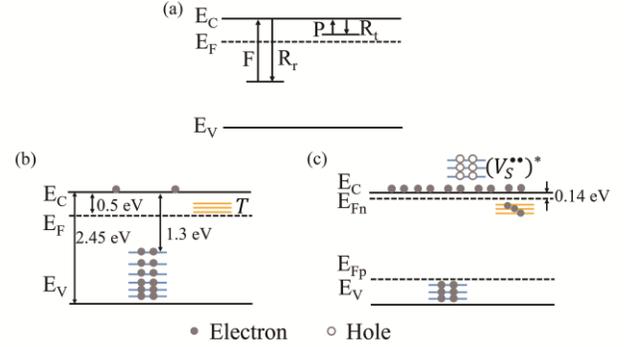

**Figure 8:** Energy transition level diagrams. (a) Standard model of photoconductivity including trap and recombination levels. (b) Energy level diagram for photoconductive CdS at equilibrium in the dark. (c) One possibility for energy level diagram for CdS under illumination.

### 4.3. Conclusions

The proposed mechanism for large and persistent photoconductivity is atomic rearrangement following a redox reaction at an optically-active defect. In both ZnO and CdS, the transition $V_{An}^\times + h\nu \to (V_{An}^{\bullet\bullet})^* + ph + 2e^-$ is accompanied by a large movement of the surrounding cations (**Fig. 1** and Supplemental Materials) [28]. With the defect in the neutral state ($V_{An}^\times$), the cations are attracted to two un-bound electrons and move inwards relative to their crystal lattice positions, into the space that would otherwise be occupied by a large anion. After the transition to the doubly-charged state ($V_{An}^{\bullet\bullet}$), the cation-cation interaction is no longer screened and the cations flee the defect site, moving outwards relative to their lattice positions. This atomic motion physically isolates the trapped minority carriers, providing a kinetic barrier to recombination.

The size of this effect in ZnO and CdS can be understood by comparing the fractional change in ion-ion distance due to redox transitions at all intrinsic defects in these materials. A comparison of density functional theory (DFT) ground-state calculations, performed by different research groups for ZnO and CdS, finds a striking quantitative agreement: the fractional change in cation-cation distance due to the anion vacancy double oxidation reaction is 29% and 33% in ZnO and CdS, respectively [23,24,28]. These are much larger than the atomic rearrangements due to redox reactions at other intrinsic defects [28]. This suggests that additional materials with long-lived, non-equilibrium, photo-excited states may be identified by analyzing ground-state calculations of atomic motion following defect redox reactions. The idea that atoms rearrange to create physical separation between photo-excited charge carriers, thus slowing the decay to equilibrium, is reminiscent of "DX-center" dopant de-activation in AlGaAs, to long photo-excitation lifetimes observed in systems with polaron excited states including organic semiconductors and halide perovskites, and to biological light absorbers such as photosystem II and dye molecules used for solar energy conversion [13–19].

Persistent photoresponse often results from spatial non-uniformity, such as potential barriers due to resistive secondary phases or high doping levels. Spatial inhomogeneity is invoked to explain apparent recombination cross section ($\sigma_{CS}$) many orders of magnitude smaller than $10^{-23}$ cm$^2$ [35–37]. For the samples studied here, the large range of decay rates measured and the systematic dependence of such on the CBD bath chemistry suggest that inhomogeneous photoconductivity is not dominant. We can estimate $\sigma_{CS}$ as $(vn\tau)^{-1}$, where $v$ is the electron thermal velocity, $n$ is the instantaneous non-equilibrium carrier concentration, and $\tau$ is an instantaneous photoconductivity decay rate. Taking $v = 10^7$ cm/s and using our measured ($n$, $\tau$) data we estimate that $\sigma_{CS}$ falls in the range $10^{-27}$ – $10^{-21}$ cm$^{-2}$ for the samples studied here. The ZWLZ model can be considered an extreme version of inhomogeneous photoconductivity, in which the potential barriers are centered around individual point defects. It is possible that individual point defects and defect clusters both contribute to LPPC in CdS, thereby linking the ZWLZ model and spatial inhomogeneity models of persistent photo-effects.

In summary, we demonstrate that the photoconductive response of CdS thin films can be widely tuned by varying the bath chemistry during CBD. We show that large photoresponse and slow decay are correlated with sulfur deficiency. Both linear and quadratic electron-hole recombination are relevant, and the rate-limiting process during photoconductivity decay is thermally-activated recombination. These observations provides experimental validation for the model that anion vacancies in II-VI semiconductors are negative-$U$ defects, exhibiting strong hole-hole correlation due to lattice distortions that respond to changes in the vacancy charge state. Directions for future research include complementary spectroscopic experimental studies and in-depth theoretical modeling of the kinetic pathways for linear and quadratic recombination.

**Acknowledgements**

to Rakesh Singh and Matthew Edwards at Arizona State University for RBS measurements.

This work was supported by an Office of Naval Research MURI through grant #N00014-17-1-2661. We are grateful